\documentstyle[aps,preprint]{revtex}
\def\lsim{\mathrel{\raise2pt\hbox to 8pt{\raise -5pt\hbox{$\sim$}\hss{$<$}}}}
\draft

\newcommand{\ba}{\begin{eqnarray}}
\newcommand{\ea}{\end{eqnarray}}

\begin{document}

\title{On the Relativistic Foundations of Pseudospin Symmetry\\ 
in Nuclei} \author{
J.N. Ginocchio$^{1,*}$ and A. Leviatan$^{2,1,\dagger}$}
\address{$^{1}$~Theoretical Division, Los Alamos National Laboratory, Los
Alamos, New Mexico 87545, USA}
\address{$^{2}$~Racah Institute of Physics, The Hebrew University,
Jerusalem 91904, Israel}

\maketitle

\begin{abstract}
We show that the generators of pseudospin symmetry are the non -
relativistic limit of the generators of an SU(2) symmetry which leaves
invariant the Dirac Hamiltonian with scalar and vector potentials equal 
in magnitude but opposite in sign, $V_V = - V_S$. Furthermore, within this
framework, we demonstrate that this symmetry may be approximately conserved
for realistic scalar and vector potentials.
\end {abstract}

\vspace{5cm}
\noindent *E-mail address: gino@t5.lanl.gov\\
$^{\dagger}$E-mail address: ami@vms.huji.ac.il\\

\pagebreak

Pseudospin doublets were introduced almost thirty years ago into nuclear
physics to accommodate an observed near degeneracy of certain normal -
parity shell-model orbitals with non - relativistic quantum numbers 
($n_r$, $\ell$,$j = \ell + 1/2)$ and 
($n_{r}-1, \ell + 2$, $j = \ell + 3/2$) 
where $n_r$, $\ell$, and $j$ are the
single-nucleon radial, orbital, and total  angular momentum quantum
numbers, respectively \cite {kth,aa}. The doublet structure is expressed 
in terms of a ``pseudo'' orbital angular momentum
$\tilde{\ell}$ = $\ell$ + 1 and ``pseudo'' spin, $\tilde s$ = 1/2. For
example, $(n_r s_{1/2},(n_r-1) d_{3/2})$ will have $\tilde{\ell}= 1$ ,
$(n_r p_{3/2},(n_r-1) f_{5/2})$ will have $\tilde{\ell}= 2$, etc. 
These doublets are almost degenerate with
respect to pseudospin, since $j = \tilde{\ell}\ \pm \tilde s$ for the
two states in the doublet.  This symmetry has been used to explain
features of deformed nuclei \cite{bohr}, including superdeformation 
\cite{dudek} and identical bands \cite {twin,ben}, and to establish an 
effective shell - model coupling scheme \cite {draayer2}. 
Therefore there is an interest in
understanding the origin of this ``symmetry'' 
\cite {bahri,draayer,gino}.  
The work of \cite{draayer} has shown that a microscopic
normal$\rightarrow$pseudo transformation leads to an effective reduction 
of the spin-orbit splitting in the single-particle energy spectra derived
from realistic (one boson exchange) internucleon potentials, in line
with relativistic mean-field estimates \cite{bahri}.
The work of \cite{gino} has shown that 
quasi-degenerate pseudospin doublets in nuclei arise
from the near equality in the magnitudes of
an attractive scalar, $V_S$, and a repulsive vector, $V_V$, relativistic
mean fields, $V_S \sim \ - V_V$, in which the nucleons move.
A near equality in the magnitude
of mean fields seems to be a universal feature of relativistic theories
ranging from relativistic field theories with interacting nucleons and
mesons \cite {wal}, to nucleons interacting via Skyrme-type interactions 
\cite{mad,madp}, to QCD sum rules \cite{furn}. Recently realistic 
relativistic mean fields were shown to exhibit this approximate 
pseudospin symmetry in both the energy spectra and wave 
functions \cite {gino2}. In this paper we identify the algebra responsible
for pseudospin symmetry as a non - relativistic limit of an SU(2)
symmetry of a Dirac Hamiltonian with $V_S = - V_V$.

The Dirac Hamiltonian, H, with an external scalar, $V_S$, and vector,
$V_V$, potentials is given by:
\ba
H = \mbox{\boldmath $\alpha\cdot p$} 
+ \beta (m  + V_S) + V_V  ~,
\label {dirac}
\ea
where we have set $\hbar = c =1$ and \mbox{\boldmath $\alpha$}, 
$\beta $ are the usual Dirac matrices \cite {mul}. The Dirac Hamiltonian 
is invariant under an SU(2) algebra for two limits: $V_S$ = $V_V$
and  $V_S = - V_V$ \cite{smith,bell,eyre}, ($V_S, V_V$ are spin
scalars). The former limit has application to the spectrum of mesons for 
which the spin - orbit splitting is small. We shall show that the latter 
limit has relevance to understanding the origin of pseudospin symmetry 
in nuclei.

The generators for the SU(2) algebra, ${\hat S}_i$, which commute with the
Dirac Hamiltonian, $[\,H\,,\, {\hat S}_i\,] = 0$, for the case when 
$V_S= -V_V$ are given by \cite{bell}
\ba
{\hat S}_i =  {\mbox{\boldmath $\alpha\cdot p$} 
\ {\hat s}_i \ \mbox{\boldmath $\alpha\cdot p$} \over p^2}
\ {(1 +\beta)\over 2} + {\hat s}_i\,  {(1 - \beta)\over 2} ~,
\label {bell}
\ea
where ${\hat s}_i = \sigma_i/2$ are the usual spin generators and
$\sigma_i$ the Pauli matrices. This reduces to
\ba
{\hat S}_i = \left ( {{\hat {\tilde s}_i} \atop 0 } 
{ 0 \atop { {\hat s}_i}}\right ) ~,
\label {gen}
\ea
where
\ba
{\hat { \tilde s}_i} =  U_p\ {\hat s}_i \ U_p = 
{2\,\mbox{\boldmath $s\cdot p$} \over p^2}\ p_i - {\hat s}_i ~.
\label {ugen}
\ea
In (\ref {ugen}) $U_p = \, {\mbox{\boldmath $\sigma\cdot p$} \over p}$ 
is the momentum-helicity unitary operator introduced in \cite {draayer} 
that accomplishes the transformation from the
normal shell model space to the pseudo shell model space while preserving
rotational, parity, time - reversal, and translational invariance. 
What is significant is
that the same unitary transformation appears in the non - relativistic
limit of the generators of a symmetry possessed by the Dirac Hamiltonian 
for $V_S = - V_V$. In this case, the eigenfunctions of the Dirac 
Hamiltonian, $H \Psi = {\cal E} \Psi$ are also doublets 
($S=1/2$, $\tilde\mu =\pm 1/2$) with respect to the SU(2) generators 
${\hat S}_i$ of Eq. (\ref{gen})
\ba
{\hat S}_z\,\Psi_{\tilde \mu} &=& \tilde\mu\, \Psi_{\tilde\mu} ~,
\nonumber\\
{\hat S}_{\pm}\,\Psi_{\tilde \mu} &=& {\sqrt{(1/2 \mp {{\tilde
\mu}})( 3/2 \pm {{\tilde \mu}}) }}\, \Psi_{{\tilde \mu} \pm 1} ~,
\qquad\qquad  (\,V_S = - V_V\,)
\label {ugen2}
\ea
where ${\hat S}_{\pm} = {\hat S}_{x} \pm i{\hat S}_y$.

In general, the eigenfunctions of the Dirac Hamiltonian, 
have an upper ($\Psi_{+}$) and lower ($\Psi_{-}$) components, 
$\Psi_{\pm} = {(1 \pm \beta)\over 2}\ \Psi$. 
In nuclear spectroscopy, the lower component, which is small \cite
{gino2}, is usually ignored.  The upper component satisfies the second
order differential equation,
\ba
H_+\, \Psi_+  = ({\cal E} - m)\, \Psi_+ ~,
\label {shr}
\ea
where
\ba
H_+ = H_+^{ps} + V_V + V_S ~,
\label {ham}
\ea
and the Hamiltonian $H_+^{ps}$ is given by
\ba
H_+^{ps} = \mbox{\boldmath $\sigma\cdot p$} 
\, {1 \over {\cal E} + m  + V_S - V_V}\,  
\mbox{\boldmath $\sigma\cdot p$} ~ .
\label {hamps}
\ea
Using the identity $\mbox{\boldmath $\sigma\cdot p$}\, 
\mbox{\boldmath $\sigma\cdot p$} = p^2$, it is easy to show that
\ba
\left [\,{\hat{ \tilde s}_i} \, , \, H_+^{ps}\, \right ] = 0 ~;
\label {com}
\ea
hence $H_+^{ps}$ conserves pseudospin.
As seen from Eq. (\ref{ugen}), the ${ \hat{ \tilde s}_i}$ are obtained
by a unitary transformation from the ordinary-spin operators 
${\hat s}_i$, hence generate an SU(2) algebra
\ba
\left [\,{\hat { \tilde s}_i}\, ,\, {\hat { \tilde s}_j}\, \right ] 
&=& i\epsilon_{ijk} \, {\hat {\tilde s}_k} ~.
\label {su2}
\ea
This means that, for $V_S = - V_V$, the non-relativistic single - nucleon
wave functions (Dirac upper components) can then be labeled by pseudospin, 
${\tilde s}= 1/2$, its projection, 
$ {{\tilde \mu}},\, {\tilde \mu} = \pm 1/2, \Psi_{+,{\tilde\mu}}$, 
and that these doublets are degenerate. The pseudospin generators connect 
wave functions of the doublets in the usual way,
\ba
{\hat{\tilde s}}_z \Psi_{+,{{\tilde \mu}}} &=& {{\tilde \mu}}
\Psi_{+,{{\tilde \mu}}} ~,
\nonumber\\
{\hat{\tilde s}}_{\pm}  \Psi_{+,{{\tilde \mu}}} &=& {\sqrt{(1/2 \mp {{\tilde
\mu}})( 3/2 \pm {{\tilde \mu}}) }}\ \Psi_{+,{{\tilde \mu} \pm 1}} ~,
\qquad\qquad  (\,V_S = - V_V\,)
\label {me}
\ea
where 
${\hat{\tilde s}}_{\pm} = {\hat{\tilde s}}_x \pm i{\hat{\tilde s}}_y$.

However, in the exact pseudospin limit, there are no bound Dirac valence
states for realistic mean fields \cite {gino} and therefore nuclei would
not exist if pseudospin symmetry were exact. 
Nevertheless, it is possible to have small pseudospin symmetry breaking 
and have the requisite number of bound Dirac
valence states \cite{gino,gino2} for nuclei to exist. 
Furthermore, we claim that the
relationship between the wave functions of the doublets given in (\ref
{me}) will still be approximately
valid. In order to prove this we see from (\ref {ugen2}), that, in the 
pseudospin symmetry limit, the lower components are ordinary spin 
doublets with spin, $s= 1/2$, and projection $\tilde\mu$, 
$\tilde\mu = \pm 1/2$:
\ba
{ {\hat s}}_z \Psi_{-,{\tilde\mu}} &=& {\tilde\mu} 
\, \Psi_{-,{\tilde\mu}} ~ ,
\nonumber\\
{ {\hat s}}_{\pm}  \Psi_{-,{\tilde\mu}} &=& 
{\sqrt{(1/2 \mp {\tilde\mu})( 3/2 \pm {\tilde\mu})}}\, 
\Psi_{-,{\tilde\mu \pm 1}} ~,
\qquad\qquad  (\,V_S = - V_V\,) ~.
\label{lme}
\ea
These spin doublets are degenerate in energy and the wave functions
$\Psi_{-,\tilde\mu},$ $\tilde\mu = \pm 1/2, $ have the same 
dependence on the spatial coordinates in the pseudospin limit. 
However, in \cite {gino2}, it was shown that, even for realistic mean 
field Dirac Hamiltonians, the lower components of doublets have almost 
identical spatial wave functions, particularly for the doublets close to 
the Fermi sea. Even though \cite {gino2} considered
only spherically symmetric scalar and vector potentials, we assume this 
to be true in general and therefore label the states with spin projection 
$\tilde\mu$ referring to the dominant spin projection in the wave function. 
We now show that the pseudospin relations between the wave functions 
given in (\ref {me}) for the upper components are
approximately valid even when there is symmetry - breaking.

From the Dirac equation, $H \Psi = {\cal E} \Psi$, we derive,
\ba
\Psi_{+,{\tilde \mu}} = {1\over ({\cal E} - m - V_S - V_V)}\,
\mbox{\boldmath $\sigma\cdot p$}\, \Psi_{-,\tilde\mu} ~.
\label {wfrel}
\ea
Assuming that the
lower component is normalized to unity, the normalization of the upper
component is then
$\langle \Psi_{+,{\tilde \mu}}|
\Psi_{+,{\tilde \mu}} \rangle = {\cal N}^{-2}=
\langle\Psi_{-,\tilde\mu}| \,  \mbox{\boldmath $\sigma\cdot p$}\, 
({1\over {\cal E} - m - V_S - V_V })^2
\,  \mbox{\boldmath $\sigma\cdot p$}\, | \Psi_{-,\tilde\mu} \rangle $. 
For the realistic Dirac equation
with $V_S\neq - V_V$, the lower component satisfies the second order 
differential equation,
\ba
\left [ \, \mbox{\boldmath $\sigma\cdot p$}\, 
{1\over ({\cal E} - m - V_S - V_V)}\, \mbox{\boldmath $\sigma\cdot p$} 
- m + V_V - V_S\, \right ] | \Psi_{-,\tilde\mu}\rangle 
= {\cal E} \, | \Psi_{-,\tilde\mu}\rangle ~.
\label {haml}
\ea
Using (\ref {haml}), we find that
\ba
{\cal N}^{-2} =  \langle \Psi_{-,\tilde\mu}| ({\cal E} -  V_V  + V_S + m)
{1\over p^2}({\cal E} -  V_V + V_S + m) | \Psi_{-,\tilde\mu} \, \rangle ~.
\label{norm}
\ea
The normalized upper component will then be
$\Phi_{+,{\tilde \mu}} = {\cal N}  \ \Psi_{+,{\tilde \mu}}$. 
Using similar manipulations, we then can
derive the matrix elements of the pseudospin operators to be,
\ba
\langle \Phi_{+,{\tilde \mu}^{\prime}} |\,  {\tilde {\hat s}_i}\,  
|\Phi_{+,{\tilde \mu}}> &=& {\cal N}^2 \langle 
{ \Psi}_{-,{\tilde\mu}^{\prime}} 
|({\cal E} -  V_V + V_S + m){{\hat s}_i\over p^2} 
({\cal E} -  V_V + V_S + m)| { \Psi}_{-,\tilde\mu} \rangle
\nonumber\\
&
\approx & \langle { \Psi}_{-,{\tilde\mu}^{\prime}} |\,  {\hat s}_i \, 
| { \Psi}_{-,\tilde\mu}\rangle ~,
\label {psme}
\ea
where the last step follows from the fact that the lower component 
has the spin and spatial parts approximately separated \cite {gino2} 
and from (\ref {norm}). Therefore, the near identity of the
spatial part of the lower component wave function for the pseudospin
doublet insures that the
pseudospin doublets are approximately connected by the pseudospin
generators even for realistic situations in which pseudospin is broken:
\ba
\langle \Phi_{+,{\tilde \mu}^{\prime}} |\, {\hat {\tilde s}}_z \, 
|\Phi_{+,{\tilde \mu}}\rangle \,
&\approx & {\tilde \mu} \, \delta _{{\tilde \mu}^{\prime}, 
\tilde\mu} ~,
\nonumber\\
\langle \Phi_{+,{{\tilde \mu}}^{\prime}} 
|{{\hat {\tilde s}}}_{\pm} \, |\Phi_{+,{{\tilde \mu}}}\rangle 
&\approx &
\sqrt{(1/2 \mp {{\tilde \mu}})( 3/2 \pm
{\tilde \mu}) }\,
\delta_{{\tilde \mu}^{\prime}, {\tilde \mu} \pm 1} ~,
\qquad\qquad (\, V_S\approx - V_V\,) ~.
\label {apprme}
\ea

In the case in which the potentials are spherically symmetric and 
satisfy $V_S = - V_V$, the Dirac Hamiltonian has an additional invariant 
SU(2) algebra; namely,
\ba
{\hat L}_i =
\left ( {\hat {\tilde \ell_i} \atop 0 } 
{ 0 \atop { {\hat \ell}_i} }\right ),
\label {jgen}
\ea
where $\hat {\tilde \ell}_i = U_p\, {\hat \ell}_i$ $U_p$ is the 
pseudo-orbital angular momentum operator, ${\hat\ell}_i$ is the 
orbital angular momentum operator, while 
${\hat j}_i = {\hat {\tilde \ell}_i} + {\hat {\tilde s}_i}
= U_p\,(\, {\hat \ell}_i + {\hat s}_i \, )\, U_p =
{\hat \ell}_i + {\hat s}_i $.
In this
limit, the Dirac wave functions are eigenfunctions of the 
Casimir operator of this algebra, 
$\mbox{\boldmath ${\hat L}\cdot {\hat L}$}\, 
|\Phi_{{\tilde \ell},j,m_j}\rangle 
= {\tilde\ell} ({\tilde \ell} + 1) |\Phi_{{\tilde \ell},j,m_j}\rangle $, 
where we have used a coupled basis,
${\vec j} = {\vec{\tilde \ell}} + {\vec{\tilde s}}$, where 
$j$ is the eigenvalue of the total
angular momentum operator ${\hat J}_i = {\hat L}_i + {\hat S}_i,\  
\mbox{\boldmath ${\hat J}\cdot{\hat J}$}\, 
|\Phi_{{\tilde \ell},j,m_j}\rangle 
= j(j + 1)|\Phi_{{\tilde\ell},j,m_j}\rangle $, and $m_j$ 
is the eigenvalue of ${\hat J}_z$. 
Thus pseudo-orbital
angular momentum as well as pseudospin are conserved in the spherical
limit and $V_S = -V_V$. (We note in passing that the operator 
${\hat K} = 
-\beta\left(\mbox{\boldmath $\sigma\cdot {\hat \ell}$} +1 \right)$
which is conserved for arbitrary spherically symmetric $V_S$ and $V_V$
potentials \cite{mul} is given by 
$\hat K = 2\,\mbox{\boldmath ${\hat L}\cdot{\hat S}$} + 1$ ).
From (\ref{jgen}), we see that the lower component wave function 
will have spherical harmonic of rank ${\tilde \ell}$ coupled to spin to 
give total angular momentum $j$. Since $\mbox{\boldmath $\sigma\cdot p$}$ 
conserves the total angular momentum but $\mbox{\boldmath $p$}$ 
changes the orbital angular momentum by one unit because of parity
conservation, Eq. (\ref {wfrel}) tells us that the the upper component 
also has total angular momentum $j$, but orbital
angular momentum $\ell = {\tilde \ell} \pm 1$. 
If $j = {\tilde \ell} + 1/2$, then it follows that
$\ell = {\tilde \ell} + 1$, whereas if $j = {\tilde \ell} - 1/2$, then
$\ell = {\tilde \ell} - 1$. 
This agrees with the pseudospin doublets originally observed 
\cite{kth,aa} and discussed at
the beginning of this paper.  However, the results here are very general
and apply to
non - spherical nuclei as well \cite {bohr,draayer3}. For example, for
axially symmetric deformed nuclei, there is a U(1) generator corresponding
to the pseudo-orbital angular momentum
projection along the symmetry axis which is conserved in addition to the
pseudospin for $V_S = - V_V$,
\ba
{\hat \lambda} = \left ( {{\hat {\tilde \Lambda}} \atop 0 } 
{ 0 \atop {{\hat \Lambda}} }\right ),
\label {Lgen}
\ea
where $\hat {\tilde \Lambda} = U_p\  \hat \Lambda\ U_p$.
In this case the Dirac wave functions are eigenfunctions of ${\hat
\lambda}$, ${\hat
\lambda}\ |\Phi_{{\tilde \Lambda},{ \Omega}}\rangle = {\tilde
\Lambda}  |\Phi_{{\tilde \Lambda},\Omega}\rangle $, where 
${\Omega}$ is the total angular
momentum projection, ${ \Omega} = {\tilde \Lambda} + {\tilde \mu}$, which
has the same value for the upper and lower components since 
$\mbox{\boldmath $\sigma\cdot p$}$ conserves the total angular
momentum projection. Thus
${\Omega} = {\tilde \Lambda} \pm 1/2$, corresponding exactly to the quantum
numbers of the pseudospin doublets
for axially deformed nuclei discussed in \cite {bohr}. Again we expect
these multiplet relationships to be approximately valid for the realistic
situation, $V_S \approx - V_V$,
following our previous discussion.

The unitary transformation, $U_p $ (\ref {ugen}), has been used to 
transform the non-relativistic spherical shell model space to the 
pseudo shell model space \cite{draayer}. 
Basically, as seen from Eq. (\ref{wfrel}) and subsequent discussion,
in the limit of $V_S = - V_V$, this
transformation transforms the normalized Dirac upper component wave
functions into the normalized Dirac lower components. 

The Hamiltonian for the upper components in the realistic case
which is given in (\ref {ham}), leads to a Schr\"odinger equation,
\ba
&&
\left [\,{1 \over ( 2m +  V_S - V_V)}\,p^2 + 
{1 \over ( 2m +  V_S - V_V)^2} \left [ \,
\mbox{\boldmath $\sigma\cdot p$}\, , \,
V_V - V_S\,\right ]\, \mbox{\boldmath $\sigma\cdot p$} 
+ V_V + V_S \right] |\Phi_{+,{\tilde \mu}}\rangle 
\nonumber\\
&&
\qquad\qquad\qquad\qquad\qquad
\qquad\qquad\qquad\qquad\qquad
= ({\cal E} - m)\ |\Phi_{+,{\tilde \mu}} \rangle ~,
\label {2nd}
\ea
where we have made the approximation that 
$ (2m +  V_S - V_V) >> m - {\cal E}$, 
the binding energy. This Schr\"odinger equation is not of the form that
follows from the usual shell model
single - nucleon Hamiltonian because it has a spatial-dependent mass 
term and, since $(V_V - V_S)/ 2m \approx 0.41$ \cite {mad} 
in the nuclear interior, the spatial 
dependence cannot be neglected.
Furthermore, the second term in (\ref {2nd}), which produces a spin - orbit
term for spherically symmetric potentials, produces a momentum-dependent 
spin term as well. Although the helicity transformation
of non-relativistic momentum-independent Hamiltonians \cite{draayer} and 
shell-model wave functions \cite{draayer4}
induces momentum dependence, the above analysis demonstrates that
the non-relativistic wave function (Dirac upper component) has momentum
dependence {\it before} the helicity transformation
since it is a solution of a momentum-dependent Hamiltonian with a spatially
dependent effective mass.

One can ask if there are tests of the pseudospin doublet
wave functions. One ingenious attempt was applied to polarized medium
energy proton scattering for which the scalar optical potential is
approximately equal and opposite in
sign to the vector optical potential
\cite {fred}. Pseudospin predictions of the spin polarization and spin
rotation functions were shown to be badly broken experimentally. 
The reason for this is as follows. In the strict symmetry limit the 
scattering functions for the partial waves depends only on 
${\tilde \ell}$, not on the pseudospin and hence are
equal for pseudospin doublets. 
When the symmetry is broken the difference in the scattering
functions for doublets is proportional to $2{\tilde \ell} + 1$, 
just like the energy splitting for
pseudospin doublets for bound orbitals
\cite {gino,gino2}.  For medium energy proton scattering large orbital
angular momentum, and hence, large pseudo - orbital angular momentum,
dominate the scattering whereas
in heavy nuclei the occupied orbitals have relatively small pseudo -
orbital angular momenta (for example, the largest for an occupied bound
orbital in $^{208}$Pb is ${\tilde\ell} = 4$). 
For this reason tests of the bound state wave functions may be
more fruitful.

In summary, we have shown that a relativistic symmetry of the Dirac
Hamiltonian with
$V_V = - V_S$ reduces to the pseudospin symmetry in the non -
relativistic limit. We have suggested that the pseudospin 
doublet wave functions of Dirac's upper components may still possess 
this symmetry even in the realistic $V_V \approx - V_S$ situation. 
These results are expected to 
apply to non - spherical as well as spherical
nuclei, in line with the validity of the pseudospin concept shown for
the oscillator shell model with arbitrary quadrupole deformations 
\cite{draayer3,draayer4}. We also conclude 
that the conventional non - relativistic shell model 
single-nucleon Hamiltonian could be made to be consistent with the
relativistic mean field, and, as a
consequence, pseudospin symmetry, by including a spatial - dependent 
mass and a momentum-dependent spin term in the mean field.

We thank A. S. Goldhaber for discussions and for informing us about
references \cite {smith,bell,eyre,fred}. This research was supported in
part by the United States Department of
Energy and in part by a grant from the Israel Science Foundation.\\

\pagebreak


\begin{thebibliography}{99}

\bibitem{kth}

K.T. Hecht and A. Adler, Nucl. Phys. {\bf A137} (1969) 129. 

\bibitem{aa}

A. Arima, M. Harvey and K. Shimizu, Phys. Lett. {\bf B30} (1969) 517. 

\bibitem{bohr}

A. Bohr, I. Hamamoto and B. R. Mottelson, Phys. Scr. {\bf 26} (1982) 
267. 

\bibitem{dudek}

J. Dudek, W. Nazarewicz, Z. Szymanski and G. A. Leander, Phys. Rev. Lett.
{\bf 59} (1987) 1405.

\bibitem{twin}

W. Nazarewicz, P. J. Twin, P. Fallon and J.D. Garrett, 
Phys. Rev. Lett. {\bf64} (1990) 1654.

\bibitem{ben}

B. Mottelson, Nucl. Phys. {\bf A522} (1991) 1. 

\bibitem{draayer2}

D. Troltenier, C. Bahri and J. P. Draayer,  
Nucl. Phys. {\bf A586} (1995) 53.

\bibitem{bahri}

C. Bahri, J.P. Draayer and S.A. Moszkowski, Phys. Rev. Lett.
{\bf 68} (1992) 2133.

\bibitem {draayer}

A. L. Blokhin, C. Bahri and J. P. Draayer, Phys. Rev. Lett.
{\bf 74} (1995) 4149.

\bibitem{gino}

J. N. Ginocchio, Phys. Rev. Lett. {\bf 78} (1997) 436. 
\bibitem{wal}

B. D. Serot and J. D. Walecka, {\it The Relativistic Nuclear Many - Body
Problem} in {\it Advances in Nuclear Physics}, edited by J. W. Negele and
E. Vogt, Vol.\
{\bf 16} (Plenum, New York, 1986).

\bibitem{mad}

B. A. Nikolaus, T. Hoch and D. G. Madland, 
Phys. Rev. {\bf C46} (1992) 1757.

\bibitem{madp}

D. G. Madland, in {\it Proceedings of the International Conference on
Nuclear Data for Science and Technology}, Trieste, Italy, May 19--24, 1997
(in press). [See Table 4, Calculation (b), and discussion.]

\bibitem{furn}

T. D. Cohen, R. J. Furnstahl and D. K. Griegel, 
Phys. Rev. Lett. {\bf 67} (1991) 961.

\bibitem{gino2}

J. N. Ginocchio and D. A. Madland, submitted to Phys. Rev. C (1997).

\bibitem{mul}
W. Greiner, B. M\"{u}ller and J. Rafelski, {\it Quantum Electrodynamics of
Strong Fields} (Springer-Verlag, New York, 1985).

\bibitem{smith}
G. B. Smith and L. J. Tassie, 
Ann. Phys. {\bf 65} (1971) 352.

\bibitem {bell}
J. S. Bell and H. Ruegg, 
Nucl. Phys. {\bf B98} (1975) 151. 

\bibitem {eyre}
G. Eyre and H. Osborn, Nucl. Phys. {\bf B116} (1976) 281. 

\bibitem {draayer3}
T. Beuschel, A. L. Blokhin and J. P. Draayer, 
Nucl. Phys. {\bf A619} (1997) 119.

\bibitem{draayer4}
A.L. Blokhin, T. Beuschel, J.P. Draayer and C. Bahri,
Nucl. Phys. {\bf A612} (1997) 163.

\bibitem {fred}
J. B. Bowlin, A. S. Goldhaber and C. Wilkin, 
Z. Phys. {\bf A331} (1988) 83.


\end{thebibliography}
\end{document}